# Time dependence of quantum oscillator excitation by electromagnetic pulses

V.A. Astapenko[1], F.B. Rosmej[1,2,3], E.V. Sakhno[1]

[1]Moscow Institute of Physics and Technology (National Research University)

[2]Sorbonne University, Faculty of Science and Engineering

[3]LULI, Ecole Polytechnique, CEA, CNRS, Laboratoire pour l'Utilisation des Lasers Intenses

**Abstract**

The paper is devoted to the theoretical investigation of time dependences of quantum oscillator excitation by electromagnetic pulses for arbitrary values of field amplitude in the pulse. We consider the harmonic oscillator without relaxation and excitation between stationary states. The general formula for excitation of quantum states as function of time is derived in terms of instant energy of associated classical oscillator in the field of electromagnetic pulse. Thus, it is established that time dependence of quantum oscillator excitation is completely determined by the energy of the associated classical oscillator at given moment of time. Using derived expression time dependence of quantum oscillator excitation is investigated in details including total excitation from ground state, excitation from excited states, total excitation probability and corresponding excitation spectra.

## 1 Introduction

A quantum harmonic oscillator is a unique physical model that can be analytically described when interacting with a perturbation of an arbitrary value outside the framework of perturbation theory [1, 2]. On the other hand, the development of the methods of generation of ultra-short laser pulses (USP) [3, 4] makes it relevant to consider the interaction of such pulses with various targets [5-10] including oscillatory systems [11, 12]. So, in a previous work of the authors [12], the excitation of a quantum oscillator for the entire duration of the USP was considered, depending on the pulse duration and the frequency detuning of its carrier frequency from the own frequency of the oscillator and arbitrary value of electric field amplitude in the pulse. The evolution of the excitation spectra and the dependences of its probability on the pulse duration with increasing field amplitude were analyzed in detail analytically and numerically.

Recently, experimental methods for real-time monitoring of photo-processes induced by USP have been successfully developed [13, 14]. In this regard, an adequate theoretical description of these phenomena is gaining relevance. This topic is devoted, for example, to temporarily monitoring the excitation of Fano resonance in a helium atom and the birth of a photoelectron on a sub-femtosecond time scale [15, 16].

The present paper is dedicated to the detail theoretical analysis of the quantum oscillator excitation by USP as a function of time for various parameters of exciting pulse outside the framework of perturbation theory.

**2 General formulas**

The time-dependent excitation probability $W_{mn}(t)$ of a quantum harmonic oscillator (QO) under the action of electromagnetic pulse (EMP) is given by formula (Appendix A):

$$W_{mn}(t) = \frac{n!}{m!} v(t)^{m-n} \exp(-v(t)) |L_n^{m-n}(v(t))|^2 \qquad (1)$$

for transition $n \to m \, [m > n]$. Here $L_n^{m-n}$ are the generalized Laguerre polynomials.

A similar expression was obtained by Schwinger [1] in the theory of quantized fields for probabilities of transitions between states with different number of photons induced by prescribed current distribution.

Husimi obtained an expression for the probability of excitation of a quantum oscillator in terms of Charlier polynomials [2].

In this paper we consider oscillator without relaxation. One can see from (1) that time dependence of the excitation probability is determined by the dimensionless function $v(t)$.

As it is shown in Appendix A function $v(t)$ is equal to

$$v(t) = \frac{\varepsilon_{clas}(t)}{\hbar \omega_0}, \qquad (2)$$

here $\varepsilon_{clas}(t)$ is the energy absorbed by classical oscillator associated with quantum one under the action of EMP at the time moment *t*. We mean by an associated classical oscillator an oscillator with the same parameter values: its own frequency, mass, and charge. Frequency $\omega_0$ is its own frequency of quantum and of associated classical oscillator with mass *M* and charge *q*.

Expression for $\varepsilon_{clas}(t)$ has the form (see Appendix B):

$$\varepsilon_{clas}(t) = \frac{q^2}{2M} \left| \int_{-\infty}^{t} dt' \, E(t') \exp(i \omega_0 t') \right|^2. \qquad (3)$$

Here $E(t)$ is electric field strength in EMP as a function of time. We suggest in this paper that $E(t \to \pm\infty) \to 0$.

One can see from (3) that instant energy of classical oscillator is determined by an incomplete Fourier transform (IFT) of the electric field of the pulse calculated at the oscillator's own frequency.

Further, for concreteness we consider EMP with Gaussian envelope:

$$E(t,\omega,\tau) = E_0 \exp\left(-\frac{t^2}{2\tau^2}\right)\cos(\omega t), \qquad (4)$$

here $\omega$, $\tau$, are carrier frequency and pulse duration, and $E_0$ is the amplitude of electric field strength of the pulse.

For Gaussian pulse (4) IFT can be presented in the form [17]:

$$FG(t,\omega',\omega,\tau) = FGw(t,\omega'+\omega,\tau) + FGw(t,\omega'-\omega,\tau), \qquad (5)$$

$$FGw(t,w,\tau) = \sqrt{\frac{\pi}{8}}\, \tau E_0 \exp(-w^2 \tau^2/2)\left\{erf\left(\frac{t}{\sqrt{2}\,\tau} - i\frac{w\tau}{\sqrt{2}}\right) + 1\right\}. \qquad (6)$$

Here $erf(z)$ is error function, $\omega'$ is current frequency of incomplete Fourier transform.

Note that in long time limit ($t \to \infty$) we have instead of (5)-(6) expression for Fourier transform of the Gaussian pulse:

$$E(\omega',\omega,\tau) = \sqrt{\pi/2}\, E_0 \tau \left\{\exp\left[-(\omega'-\omega)^2 \tau^2/2\right] + \exp\left[-(\omega'+\omega)^2 \tau^2/2\right]\right\} \quad (7)$$

For near resonance case $|\omega - \omega_0| \ll \omega_0$ and $t \gg \tau$ formula (3) takes the form:

$$\varepsilon_{clas}(t \gg \tau) \cong \frac{\pi}{4}\frac{q^2 E_0^2 \tau^2}{M}\exp\left[-(\omega-\omega_0)^2 \tau^2\right]. \qquad (8)$$

Substituting (8) into (2) and the result in (1) we come to the expression for asymptotic (for $t \gg \tau$) excitation probability of QO, which was investigated in detail in the paper [12].

Fig.1 demonstrates the time dependence of the energy absorbed by associated classical oscillator (2) normalized on energy quant ($\hbar\omega_0$) for different relative de-tunings of pulse carrier frequency from the oscillator's own frequency $\delta = (\omega - \omega_0)/\omega_0$. It is seen that with an increase in detuning $\delta$, a maximum appears in the dependence $\nu(t)$, while the magnitude of this function decreases.

Another feature is presence of oscillations in time dependence of absorbed energy. The frequency of these oscillations is equal to the quantum oscillator's own frequency.

Mathematically, the presence of oscillations follows from formulas (5-6), which determine the incomplete Fourier transform of the electric field strength in a pulse. As it is seen from these formulas, oscillations that occur due to the interference between two contributions into IFT come from terms with sum and difference of its own frequencies and carrier frequencies. So they have another nature than Rabi oscillations.

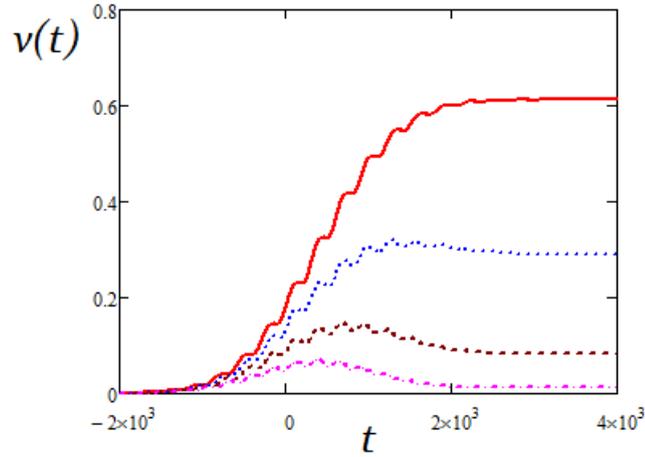

Figure 1. Time dependence of normalized energy of associated classical oscillator (2) for various spectral de-tunings: solid line - δ=0.05, dotted line - δ=0.1, dashed line - δ=0.15, dotted-dashed line - δ=0.2; $\tau=10^3$, $E_0=10^{-2}$, q=1, $M=10^4$, $\omega_0=0.01$

For the excitation of the QO from the ground state (n=0→m), expression (1) is simplified to the form

$$W_{m0}(t) = \frac{v(t)^m}{m!} \exp(-v(t)) \qquad (9)$$

It is easy to obtain from (9) the formula for total excitation probability of QO from the ground state:

$$W_{tot}(t) = \sum_{m=1}^{\infty} W_{m0}(t) = 1 - \exp(-v(t)). \qquad (10)$$

As it follows from Fig.1, function $W_{tot}(t)$ has different behavior depending upon the value of parameter δ (see details in Fig.5).

# 3 Results and discussion

The results of calculations of the time dependence of probability of QO excitation using obtained formulas are shown in the figures below for $\omega_0=0.01$, $q=1$, $M=10^4$ and different values of $E_0$, $\tau$ and $\delta$.

Fig. 2 shows the time dependence of the probability of excitation of QO from the ground to the nearest excited states ($m=1, 2, 3$) for different magnitudes of the amplitude of the electric field in the EMP. The asymptotic values of the excitation probability at large times, calculated using formulas (2, 8, 9), are also given here by thin solid lines.

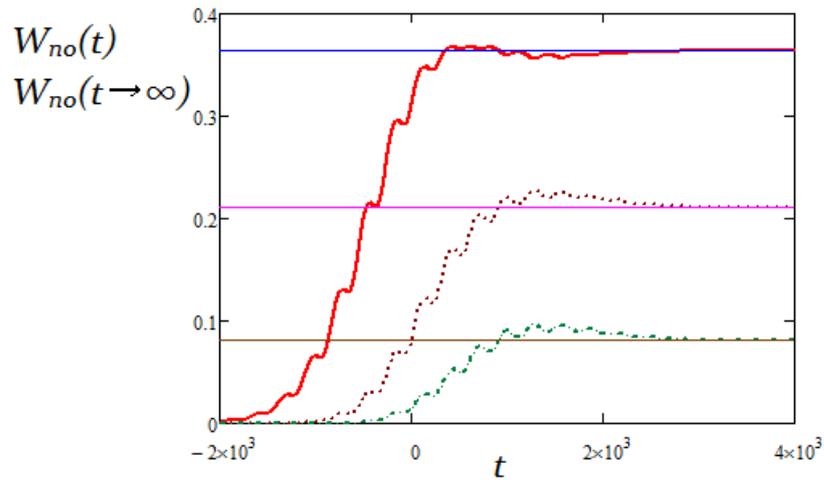

(a)

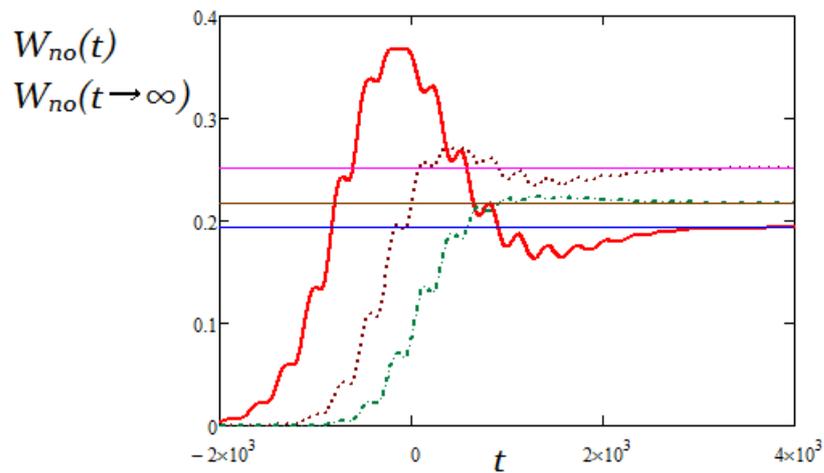

(b)

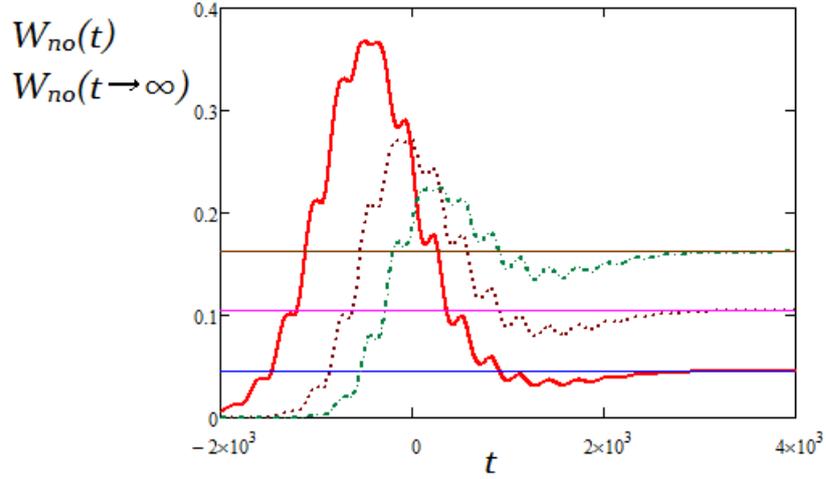

(c)

Figure 2. Time dependence of QO excitation probability from ground state for $E_0=2\cdot 10^{-2}$ (a), $E_0=3\cdot 10^{-2}$ (b), $E_0=4\cdot 10^{-2}$ (c) and different upper states: thick solid line – m=1, dotted line – m=2, dashed line – m=3, thin solid lines correspond to asymptotic values of probability calculated using (8)-(9); relative spectral detuning $\delta=0.1$

It follows from Fig.2 that, with an increase in the electric field strength, the dependence $W_{m0}(t)$ from a monotonously increasing function turns into a function with a maximum, the amplitude of which practically does not change with an increase in $E_0$ and whose position shifts slightly to the region of shortened times.

Oscillations in curves depicted in Fig.2 arise due to oscillations in the time dependence of absorbed energy (see Fig. 1) and originate because of the interference effect described above.

Fig. 3 presents the time dependence of excitation probability of 0→2 transition for different pulse durations and various spectral detuning δ.

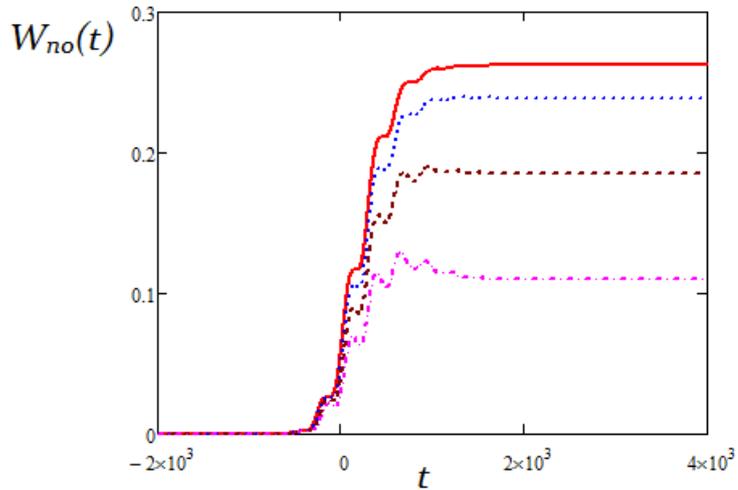

(a)

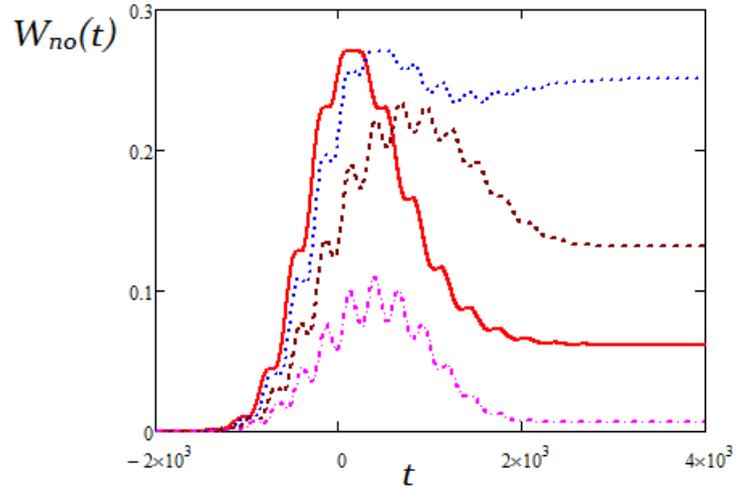

(b)

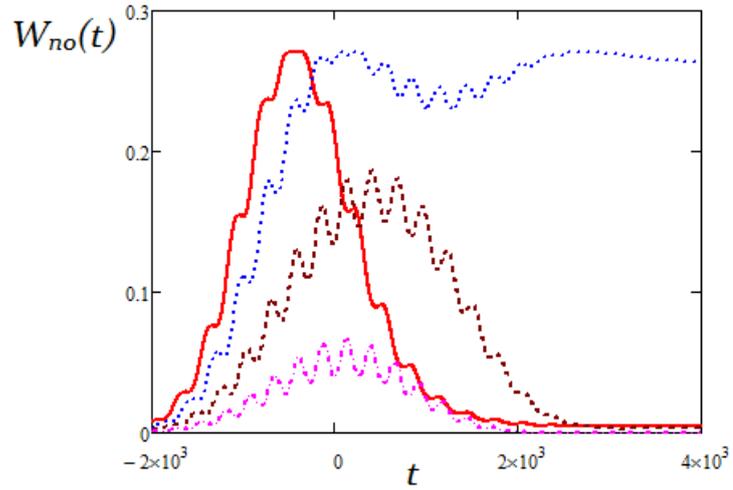

(c)

Figure 3. Time dependence of QO excitation probability on transition 0→2 for various pulse durations τ=500 (a); τ=1000 (b), τ=1500 and different de-tunings: solid line - δ=0.05, dotted line - δ=0.1, dashed line - δ=0.15, dotted-dashed line - δ=0.2; $E_0=3 \cdot 10^{-2}$

From Fig.3(a) one can see that for small pulse duration excitation probability increases almost monotonously, reaching saturation value at times $t \geq 2\tau$. In this case, the value of the stationary probability of excitation decreases with an increase of spectral detuning δ. This situation is characteristic for relatively small parameters of $v$ when $\exp(-v) \approx 1$ and perturbation theory approach is valid. With an increase in the pulse duration, the role of nonlinearity in the interaction of EMP with QO increases and the dependence of the excitation probability on time becomes more complex, as can be seen from Fig. 3 (b, c). One of the characteristic features of the time dependence in the case of longer pulses is the fact that, over time, the probability of excitation by pulses with a larger frequency detuning becomes greater than for pulses with a smaller parameter $\delta$.

The asymptotic values of the excitation probability also significantly depend on the magnitude of the detuning and the duration of the pulse. So, for example, in the case shown in Fig. 3, the maximum asymptotic value of the excitation probability corresponds to the spectral detuning $\delta=0.1$. This agrees with the dependence presented in Fig.4 which shows the asymptotic excitation probability as a function of pulse duration for different spectral de-tunings calculated using formulas (2, 8, 9). One can see from this figure that for given parameters maximum magnitude has a curve with $\delta=0.1$ in the range $\tau>600$. Note that asymptotic excitation probability was investigated in detail in paper [12], where analytical treatment was particularly given to excitation probability dependence upon pulse duration for different values of other pulse parameters (see Fig.4 from cited paper).

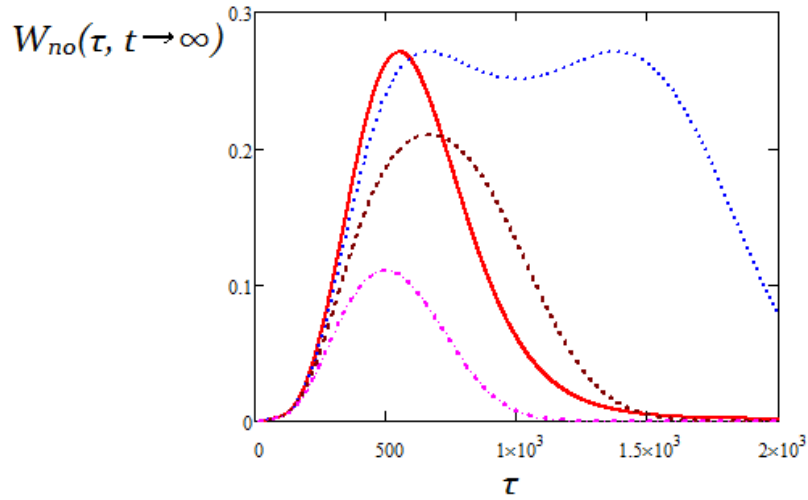

Figure 4. Asymptotic excitation probability of QO at transition 0→2 as a function of pulse duration for various detuning values: solid line - $\delta=0.05$, dotted line - $\delta=0.1$, dashed line - $\delta=0.15$, dotted-dashed line - $\delta=0.2$; $E_0=3\cdot 10^{-2}$

Total excitation probability of QO from ground state as a function of time is shown in Fig.5 (a, b, c) for various $\delta$ and $\tau$; $E_0=3\cdot 10^{-2}$. For comparison, this figure also presents the time dependence of the probability of excitation of the 0→1 transition for a pulse duration $\tau=1500$ and various values of detuning $\delta$.

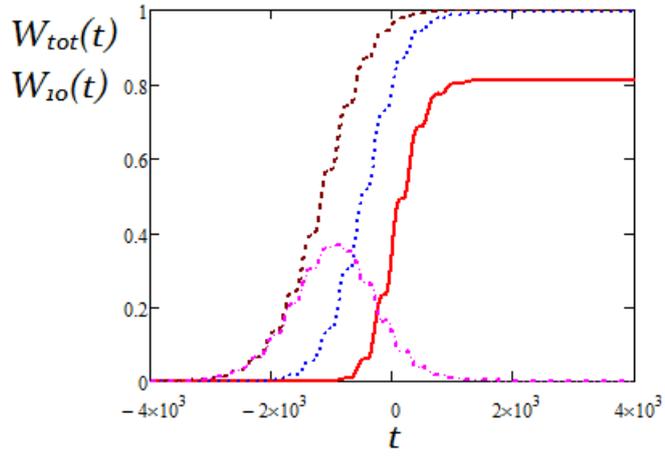

(a)

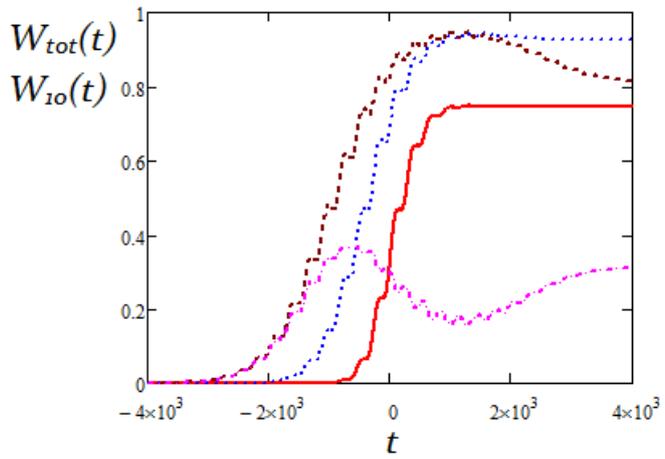

(b)

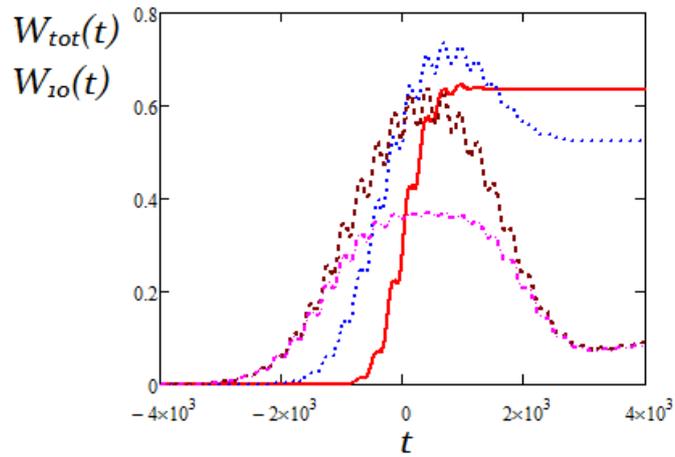

(c)

Figure 5. Total excitation probability of QO from ground state as a function of time for various de-tunings $\delta=0.05$ (a), $\delta=0.1$ (b), $\delta=0.15$ and different pulse durations: solid line - $\tau=500$, dotted line - $\tau=1000$, dashed line - $\tau=1500$; dotted-dashed line describes the excitation of transition $0\rightarrow1$; $E_0=3\cdot10^{-2}$

It is seen that in the case of a short pulse, the time dependence of the total probability of excitation of the QO does not practically change with increasing spectral detuning. On the contrary, for a long pulse, the shape of the curve changes significantly, which is clearly seen in Fig. 5(c).

Another conclusion following from Fig.5 is that the time dependence of the total probability of excitation of QO approaches its analogue for excitation of the 0→1 transition with increasing spectral detuning $\delta$.

Fig.6 demonstrates time evolution of the QO excitation spectrum for 0→1 transition (a) and for total excitation from ground state (b). The excitation spectrum in the paper refers to the dependence of the excitation probability on the relative frequency detuning $\delta = (\omega - \omega_0)/\omega_0$.

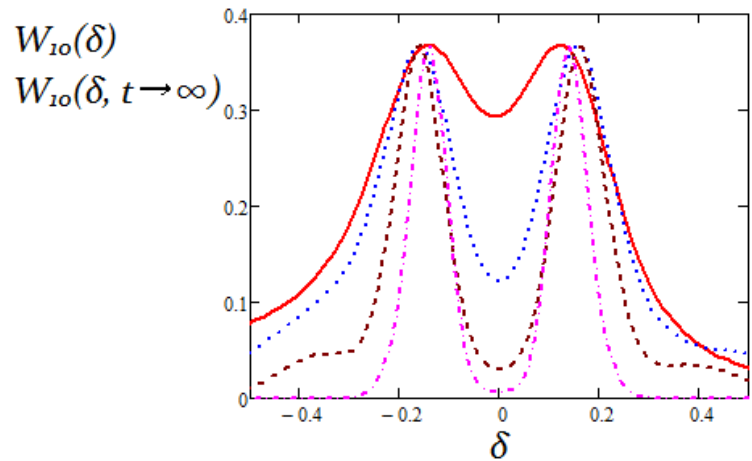

(a)

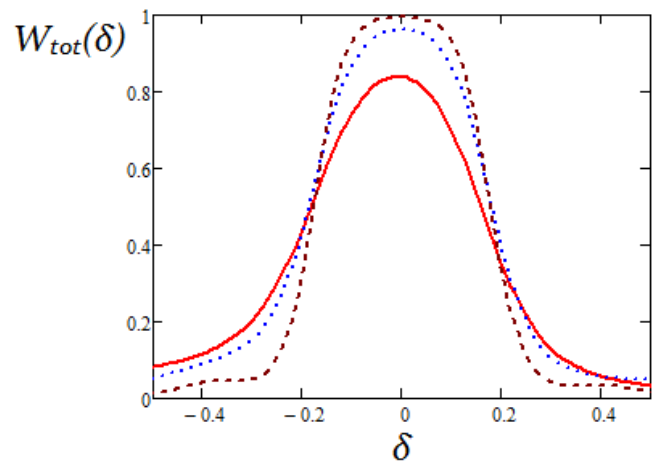

(b)

Figure 6. Time evolution of the excitation probability spectrum for 0→1 transition (a) and for total excitation of QO from ground state (b): solid line – t=10, dotted line – t=500, dashed line – t=$10^3$; $E_0$=3·$10^{-2}$, τ=1000; dotted-dashed line in Fig. 6(a) shows asymptotic spectrum for t→∞

As follows from Fig. 6 (a) the excitation spectrum of the transition 0→1 for a given field amplitude has a dip in the center at δ = 0, which, over time, will drop down to the zero value. In addition, the spectrum has a weakly expressed asymmetry, which disappears in the limit of large times. The calculation shows that for smaller values of the field amplitude, the above dip is absent for all times. The excitation spectrum in this case is a bell-shaped curve with a maximum at δ=0. The probability spectrum of the total excitation of QO has one maximum at δ = 0 for all values of the field strength in the pulse. With increase of time, the shape of this maximum changes from a bell-shaped to table-shaped curve as it can be seen from Fig. 6(b).

The described features are associated with the presence of a factor $\exp(-v)$ in the expression for probability of QO excitation. Physically, the appearance of a spectral dip in the transition between stationary states corresponds to the saturation of this transition under the action of EMP.

Let us consider now excitation from excited states of QO using formulas (A26, A30) which are follows from (1).

$$W_{mn}(t) = n!\, m!\, v(t)^{m+n}\, exp(-v(t)) \left| \sum_{k=0}^{\min(n,m)} \frac{(-1)^k\, v(t)^{-k}}{k!(m-k)!(n-k)!} \right|^2 \quad (11)$$

Parameter ν is determined by equality (2).
It should be noted that in contrast to (1), expression (11) presents the result in explicit form as a function of parameter ν and numbers of stationary states (m, n).

The results of calculations of excitation probability for transitions $n \to 3$ (n=0, 1, 2) for different values of EMP amplitude $E_0$ are presented in Fig. 7.

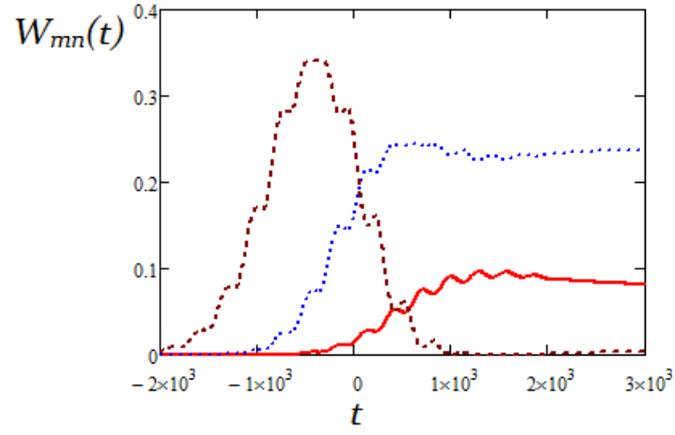

(a)

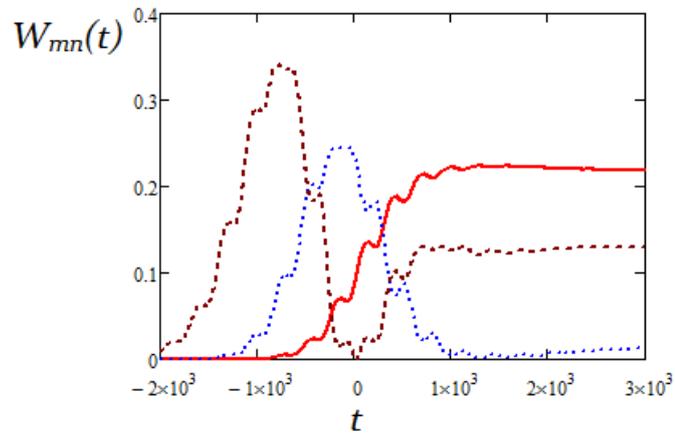

(b)

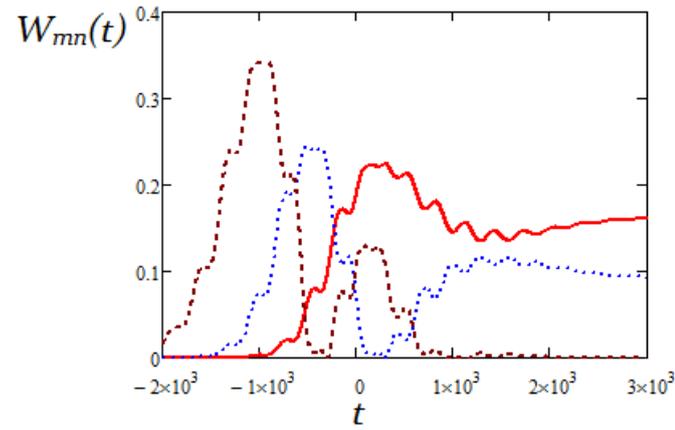

(c)

Fig.7. Time dependence of QO excitation probability by EMP with different field amplitude $E_0=2\cdot10^{-2}$ (a), $E_0=3\cdot10^{-2}$ (b) and $E_0=4\cdot10^{-2}$ (c) for transitions $n \to 3$: solid line - $0\to3$; dotted line - $1\to3$, dashed line - $2\to3$; $\delta=0.1$; $\tau=10^3$

One can see that the time dependence of excitation probability changes most strongly for the transition between the nearest energy levels (transition 2→3) so that a new maximum appears while the time dependence of excitation probability for transition 0→3 changes relatively weakly. Note that for transitions from the ground state of QO, for sufficiently strong fields, there is only one maximum in the probability of excitation as a function of time (see Fig. 2b, c).

Time evolution of excitation spectrum on 1→3 transition for different values of EMP amplitude is shown in Fig. 8.

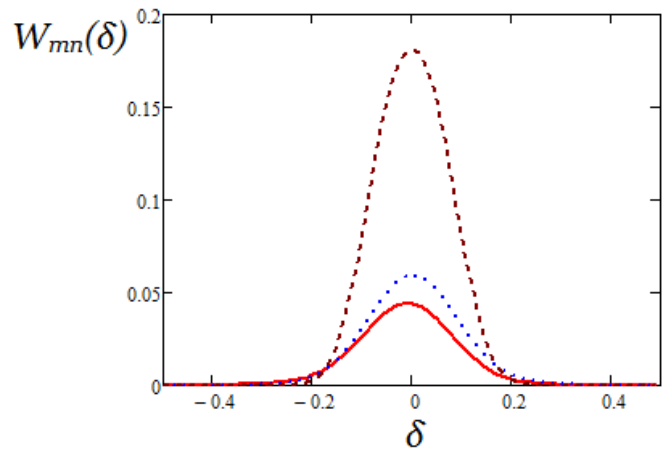

(a)

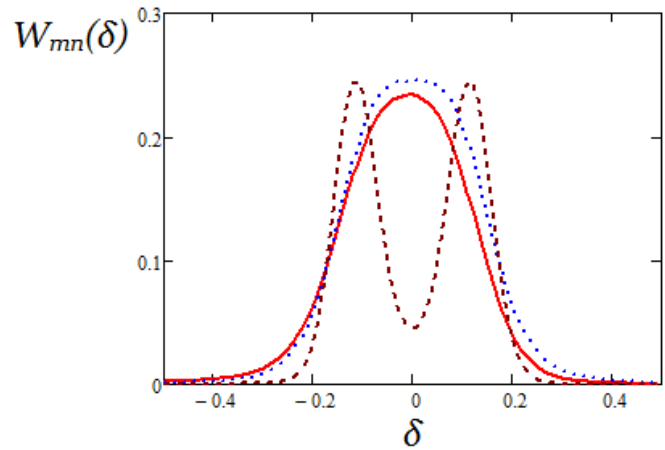

(b)

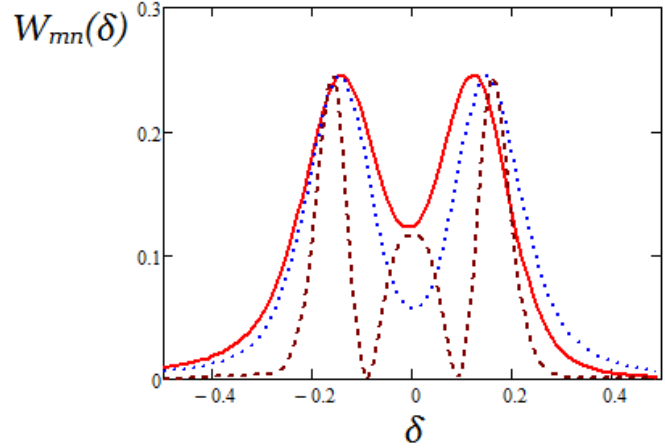

(c)

Figure 8. Time evolution of excitation spectrum on 1→3 transition of QO for different values of field amplitude $E_0=10^{-2}$ (a), $E_0=2\cdot10^{-2}$ (b), $E_0=3\cdot10^{-2}$ (c): solid line – t=10, dotted line – t=100, dashed line – t=$10^3$; pulse duration $\tau=10^3$

It can be seen from the above figure that with increasing field amplitude, the excitation spectrum becomes more and more complex. Instead of one maximum at small values of the amplitude Fig.8 (a), two maxima appear with a dip in the center Fig.8 (b), which with a further increase in the amplitude turns into a maximum Fig.8 (c). Thus, the excitation spectrum expands with increasing field, while the number of maxima and minima increases. It is significant that, at large field amplitudes, the dependence of the excitation spectrum changes more strongly with time.

It is noteworthy that the value of the probability of excitation at the maxima remains unchanged at a sufficiently large field amplitude and equal for given transition to the same value $W_{max}(1\rightarrow 3)\cong 0.245$.

## 4 Conclusion

In the paper the time dependence of QO excitation by EMP was investigated in detail numerically outside the framework of perturbation theory. We showed explicitly that this dependence is determined by the energy of the associated classical oscillator at given moment of time.

We considered both the excitation of $n\rightarrow m$ transition $W_{mn}(t)$ and the total excitation from ground state into any excited states $W_{tot}(t)$. In addition, the evolution in time of the spectrum of excitation of QO was calculated and analyzed over time in both cases.

It was shown that the behavior of function $W_{mn}(t)$ strongly depends upon EMP parameters: detuning of carrier frequency from own frequency of QO, pulse duration and amplitude of electric field strength. The analysis showed that with an increase in the mentioned values, the probability of QO excitation monotonically increasing with time turns into a function that has an envelope with a maximum and oscillations with frequency equal to its own frequency. Amplitude of these oscillations increases with increasing pulse duration $\tau$ and frequency detuning of the carrier frequency from the own frequency $\delta$. These oscillations are of a different nature than Rabi oscillations. They arise as a result of interference of contributions to the process from the positive and negative frequency components of the electric field in the pulse.

The total probability of excitation of QO in the case of short pulses remains a monotonically increasing function of time with increasing detuning $\delta$, while for long pulses the probability of excitation acquires a maximum with an increase of $\delta$.

We investigated the time evolution of the excitation spectra of QO for different amplitudes of electric field of EMP. In particular, it was shown that with an increase in the field amplitude on the spectral curve, instead of one maximum for $\delta=0$, lateral maxima appear, the position and shape of which change over time. In the case of excitation from excited state the central spectral minimum turns into a maximum over time (Fig.8(c)).The spectrum of the total excitation of QO from the ground state in the limit of strong fields is described by a table-shaped curve, the width of which decreases with time.

Finally we derived the expression of QO excitation on n→m transition (11) with explicit dependence upon numbers of states (m, n). Particularly it follows from this expression that the probabilities of mutually inverse transitions in a quantum oscillator under EMP influence are equal to each other in the prescribed field approximation.

**Acknowledgements** This research was supported by Moscow Institute of Physics and Technology in the framework of 5-top-100 program (The project No. 075-02-2019-967).

# Appendix A

## Derivation of main formulas (1-2)

We suggest that prior to excitation $(t \to -\infty)$ the oscillator was in the $n$th stationary state with a wave function (without a time factor) [18]:

$$\psi_n^{(0)}(x) = \left(\frac{M\omega}{\pi\hbar}\right)^{\frac{1}{4}} \frac{1}{\sqrt{2^n n!}} \exp\left(-\frac{M\omega}{2\hbar}x^2\right) H_n\left(x\sqrt{\frac{M\omega}{\hbar}}\right), \quad (A.1)$$

here $H_n$ – Hermite polynomials. Let's introduce dimensionless variables:

$$\tilde{x} = x\sqrt{\frac{M\omega}{\hbar}} \quad (A.2)$$

and a dimensionless function:

$$\tilde{\psi} = \left(\frac{\hbar}{M\omega}\right)^{\frac{1}{4}} \psi \quad (A.3)$$

Because of EMP action the initial function transforms to $\tilde{\psi}_n^{(0)}(x) \to \tilde{\psi}_n(x,t)$. Then, for the probability of transition between stationary states of a quantum oscillator $n \to m$ we have (at the time moment $t$):

$$W_{mn}(t) = \left|\int_{-\infty}^{+\infty} d\tilde{x}\, \tilde{\psi}_n(\tilde{x},t) \tilde{\psi}_m^{(0)}(\tilde{x})\right|^2 \quad (m > n) \quad (A.4)$$

here the function is entered [2]

$$\tilde{\psi}_n(\tilde{x},t) = \exp(i\varphi(t)) \tilde{\psi}_n^{(0)}(\tilde{x} - \tilde{\eta}(t)) \exp(i\dot{\tilde{\eta}}\tilde{x}), \quad (A.5)$$

which is an analytical solution of the temporal Schrödinger equation for a one-dimensional harmonic oscillator, which is affected by an external time-dependent force.

$$\tilde{\eta}(t) = \sqrt{\frac{M\omega}{\hbar}} \eta(t), \quad (A.6)$$

$\eta(t)$ is solution of the equation for forced oscillations of a classical oscillator under the action of an external force $f(t)$:

$$\ddot{\eta} + \omega^2 \eta = \frac{f(t)}{M}, \quad (A.7)$$

with initial conditions: $\eta(-\infty) = \dot{\eta}(-\infty) = 0$,

$$\dot{\tilde{\eta}} = \frac{d\tilde{\eta}}{\omega dt} = \frac{d\tilde{\eta}}{d\tilde{t}} \tag{A.8}$$

$$\tilde{\eta} = \sqrt{\frac{M}{\hbar\omega}}\,\eta. \tag{A.9}$$

Then we have

$$W_{mn}(t) = \frac{1}{\pi 2^n n!\, 2^m m!} \times$$

$$\times \left| \int_{-\infty}^{+\infty} d\tilde{x} \exp\left(-\frac{(\tilde{x}-\tilde{\eta})^2}{2}\right) H_n(\tilde{x}-\tilde{\eta}(t)) \exp(i\dot{\tilde{\eta}}\tilde{x}) H_m(\tilde{x}) \exp\left(-\frac{\tilde{x}^2}{2}\right) \right|^2 \tag{A.10}$$

It is necessary to calculate the integral:

$$I_1 = \int_{-\infty}^{+\infty} d\tilde{x} \exp\left(-(\tilde{x}-\tilde{\eta})^2/2\right) H_n(\tilde{x}-\tilde{\eta}) \exp(i\dot{\tilde{\eta}}\tilde{x}) H_m(\tilde{x}) \exp(-\tilde{x}^2/2). \tag{A.11}$$

Let be $\tilde{x} \equiv u$, then

$$I_1 = \int_{-\infty}^{+\infty} du \exp\left[-u^2 + u\tilde{\eta} - \frac{\tilde{\eta}^2}{2} + i\dot{\tilde{\eta}}u\right] H_m(u) H_n(u-\tilde{\eta}) \tag{A.12}$$

We introduce

$$y = \frac{i}{2}\dot{\tilde{\eta}} + \frac{1}{2}\tilde{\eta}, \tag{A.13}$$

then

$$I_1 = \exp\left(\frac{i}{2}\dot{\tilde{\eta}}\tilde{\eta} - \tilde{\eta}^2/4 - \dot{\tilde{\eta}}^2/4\right) I_2(\tilde{\eta}, y) \tag{A.14}$$

$$I_2(\tilde{\eta}, y) \equiv \int_{-\infty}^{+\infty} du\, e^{-(u-y)^2} H_m(u) H_n(u-\tilde{\eta}) = (u = x+y) =$$

$$= \int_{-\infty}^{+\infty} dx\, e^{-x^2} H_m(x+y) H_n(x+y-\tilde{\eta}) \tag{A.15}$$

We use the table integral (7.378) [19]

$$\int_{-\infty}^{+\infty} e^{-x^2} H_m(x+w) H_n(x+z) dx = 2^m \sqrt{\pi}\, n!\, z^{m-n} L_n^{m-n}(-2wz)\, [m \geq n], \tag{A.16}$$

here $L_n^{m-n}$ – generalized Lagerra polynomial.

To reduce $I_2(\tilde{\eta}, y)$ to the above table integral, we need to make a substitution:

$$\begin{cases} w = y \\ z = y - \tilde{\eta} \end{cases} \quad (A.17)$$

Then we obtain

$$I_2(\tilde{\eta}, y) = 2^m \sqrt{\pi}\, n! (y - \tilde{\eta})^{m-n} L_n^{m-n}(-2y(y - \tilde{\eta})) \quad (A.18)$$

Let's introduce the notation: $\arg = -2y(y - \tilde{\eta})$. Using (A.17) we have:

$$\begin{aligned}\arg &= -2\left(\frac{i}{2}\dot{\tilde{\eta}} + \frac{1}{2}\tilde{\eta}\right)\left(\frac{i}{2}\dot{\tilde{\eta}} + \frac{1}{2}\tilde{\eta} - \tilde{\eta}\right) = \\ &= -2\left(\frac{i}{2}\dot{\tilde{\eta}} + \frac{1}{2}\tilde{\eta}\right)\left(\frac{i}{2}\dot{\tilde{\eta}} - \frac{1}{2}\tilde{\eta}\right) = \frac{\dot{\tilde{\eta}}^2 + \tilde{\eta}^2}{2}\end{aligned} \quad (A.19)$$

or in dimension variables:

$$\arg = \frac{M\dot{\eta}^2 + \omega^2 M\eta^2}{2\hbar\omega} \equiv \frac{\varepsilon(t)}{\hbar\omega} = \frac{A(t)}{\hbar\omega} = v(t) \quad (A.20)$$

$$v(t) = \frac{A(t)}{\hbar\omega}, \quad (A.21)$$

here $A(t)$ is the work which was done under classical oscillator $\eta(t)$ by EMP. Given that

$$|y - \tilde{\eta}|^2 = \frac{\dot{\tilde{\eta}}^2 + \tilde{\eta}^2}{4} = \frac{M(\dot{\eta}^2 + \omega^2\eta^2)}{4\hbar\omega} = \frac{v}{2} \quad (A.22)$$

we find

$$|I_2|^2 = 2^{2m}\pi(n!)^2 \frac{v^{m-n}}{2^{m-n}}|L_n^{m-n}(v)|^2 = 2^{n+m}\pi(n!)^2 v^{m-n}|L_n^{m-n}(v)|^2 \quad (A.23)$$

Since

$$|I_1|^2 = \exp(-v)|I_2|^2 \quad (A.24)$$

we obtain

$$W_{mn}(t) = \frac{1}{\pi 2^n n! 2^m m!}|I_1|^2 \quad (A.25)$$

Finally find excitation probability

$$W_{mn}(t) = \tilde{W}_{mn}[v(t)] = \frac{n!}{m!}v(t)^{m-n}\exp(-v(t))|L_n^{m-n}(v(t))|^2 \quad (A.26)$$

for transition: $n \to m \ [m > n]$.

In calculations we use the following representation of generalized Lagerra polynomial [19]:

$$L_n^\alpha(x) = \frac{1}{n!} e^x x^{-\alpha} \frac{d^n}{dx^n}\left(e^{-x} x^{n+\alpha}\right). \tag{A.27}$$

Substitution (A.27) in (A.26) gives

$$\tilde{W}_{mn}(v) = \frac{1}{n!m!} e^v v^{n-m} \left|\frac{d^n}{dv^n}\left(e^{-v} v^m\right)\right|^2 \tag{A.28}$$

Note that

$$v^{n-m}\left|\frac{d^n}{dv^n}\left(e^{-v} v^m\right)\right|^2 = v^{m+n} e^{-2v} \left|\sum_{k=0}^{\min(n,m)} \frac{n!m!(-1)^k v^{-k}}{k!(m-k)!(n-k)!}\right|^2 \tag{A.29}$$

Finally we have:

$$\tilde{W}_{mn}(v) = n!m! v^{m+n} e^{-v} \left|\sum_{k=0}^{\min(n,m)} \frac{(-1)^k v^{-k}}{k!(m-k)!(n-k)!}\right|^2. \tag{A.30}$$

From (A.30) in particular follows

$$\tilde{W}_{mn}(v) = \tilde{W}_{nm}(v). \tag{A.31}$$

Thus time dependence of absorption on transition $n \to m$ ($m > n$) coincides with induced emission on reverse transition $m \to n$.

# Appendix B

For the EMP energy absorbed by the charged classical oscillator by the instant of time $t$, we have:

$$\varepsilon_{clas}(t) = q \int_{-\infty}^{t} \dot{x}(t') E(t') dt'. \tag{B.1}$$

The solution of the equation for the harmonic oscillator looks as follows (for example, see [20]):

$$x(t) = \frac{q}{M} \int_{-\infty}^{+\infty} \frac{E(\omega') \exp(-i\omega't)}{\omega_0^2 - \omega'^2 - 2i\gamma\omega'} \frac{d\omega'}{2\pi}. \tag{B.2}$$

From the formula (B.1), the expression follows:

$$\dot{x}(t) = -i\frac{q}{M}\int_{-\infty}^{+\infty}\frac{\omega' E(\omega')\exp(-i\omega' t)}{\omega_0^2 - \omega'^2 - 2i\gamma\omega'}\frac{d\omega'}{2\pi}. \qquad (B.3)$$

Substituting the formula (B.3) in (B.1), we obtain:

$$\varepsilon_{clas}(t) = \frac{q^2}{M}\int_{-\infty}^{t} dt' E(t')\int_{-\infty}^{\infty}\frac{d\omega'}{2\pi} E(\omega')\frac{i\omega'\exp(-i\omega' t')}{\omega'^2 - \omega_0^2 + 2i\omega'\gamma}. \qquad (B.4)$$

The frequency integral in the right-hand side of the equation (B.4) can be calculated with the use of the residue theorem. In the limit $\gamma \to 0$ this integral is equal to

$$\int_{-\infty}^{\infty}\frac{d\omega'}{2\pi}\frac{i\omega'\exp(-i\omega'(t'-t''))}{\omega'^2 - \omega_0^2 - 2i\omega'\gamma} = \theta(t'-t'')\cos[\omega_0(t'-t'')], \qquad (B.5)$$

where $\theta(\tau)$ is the Heaviside theta function.

Substituting the right-hand side of the equation (B.5) in the formula (B.4), we find:

$$\varepsilon_{clas}(t) = \frac{q^2}{M}\int_{-\infty}^{t} dt' E(t')\int_{-\infty}^{t'} dt'' E(t'')\cos[\omega_0(t'-t'')] =$$

$$= \frac{q^2}{2M}\int_{-\infty}^{t} dt' E(t')\int_{-\infty}^{t} dt'' E(t'')\cos[\omega_0(t'-t'')] = \qquad (B.6)$$

$$= \frac{q^2}{2M}\left|\int_{-\infty}^{t} dt' E(t')\exp(i\omega_0 t')\right|^2$$

In going to the second equality in (B.6), the fact was used that the integrand is symmetric relative to the rearrangement of integration variables: $t' \leftrightarrow t''$.